\documentclass[%
 reprint,
 amsmath,amssymb,
 aps,
]{revtex4-2}

\usepackage{color}
\usepackage{graphicx}
\usepackage{dcolumn}
\usepackage{bm}

\begin{document}

\preprint{APS/123-QED}

\title{Modeling the time evolution of a camphor rotor perturbed by a stationary camphor source.
 }

\author{Jerzy Gorecki}
 \email{jgorecki@ichf.edu.pl}
 \affiliation{Institute of Physical Chemistry, Polish Academy of Sciences, 01-224 Warsaw, Poland}%

\author{Yuki Koyano}
\affiliation{Graduate School of Human Development and Environment, Kobe University, Kobe 657-0011, Japan
}%
\author{Hiroyuki Kitahata}
\affiliation{Department of Physics, Graduate School of Science, Chiba University, Chiba 263-8522, Japan
}%

\date{\today}

\begin{abstract}
A self-propelled motion resulting from the dissipation of camphor molecules on the water surface has been attracting scientific attention for more than 200 years. A generally accepted description of the phenomenon includes equations for the object motion coupled with the hydrodynamics of Marangoni flows and the time evolution of camphor surface concentration. The solution of such equations is a numerically complex problem. In recent publications, an alternative approach based on Hamiltonian including the potential term representing Marangoni interactions has been applied to simulate the time evolution of camphor rotors. Such a model represents a significant numerical simplification if compared to the standard description. Here, we comment on the applicability of Hamiltonian approach by applying it to a single camphor rotor perturbed by a camphor disk fixed on the water surface. We demonstrate that in such a case, the approach leads to the results qualitatively different from the experimental ones. Therefore, we doubt in its applicability to describe the time evolution of interacting camphor rotors. We also show that the approximation of the Marangoni forces by a potential gives more realistic results if used together with the equation of motion that includes the hydrodynamic friction. Still, a better agreement with experiments can be obtained by considering an additional equation for the time evolution of camphor surface concentration.

\end{abstract}

\maketitle

\section{Introduction}
The complex self-propelled motion of camphor grains on a water surface attracted attention centuries ago \cite{doi:10.1098/rspl.1889.0099,doi:10.1098/rspl.1860.0124,Skey1878}. This phenomenon represents an interesting scientific problem because it shows the direct conversion of chemical energy into mechanical energy. Thermodynamically, a self-propelled camphor grain on the water surface is an open system in which camphor molecules migrate from the source, form a surface layer, evaporate from it, and finally scatter in the atmosphere. The chemical energy of camphor is transformed into mechanical energy of a moving piece, and then it is dissipated as heat by hydrodynamic drag forces. The self-propelled motion of a camphor piece can be explained by the decrease in surface tension of water with an increase in camphor surface concentration. The piece is pushed by the differences in camphor surface concentration around its periphery that can generate a non-zero force (or torque) \cite{doi:10.1021/la970196p,C5CP00541H}. There are self-propelled objects like camphor boats \cite{Kohira_Langmuir,Shimokawa_PhysRevE.98.022606,Fujinami_boat_doi:10.1246/cl.140201,alcohol_boat} for which the symmetry of surface concentration of active molecules is intentionally broken to ensure a sustained driving force. If a camphor source is symmetrical, then the motion can be initiated by fluctuations in camphor surface concentration around the object. After the motion starts, the symmetry is broken, and the initially selected direction of motion is supported up to the moment when a strong perturbation, like a collision with container walls, occurs.   

The standard theoretical description of self-propelled objects combines equations for the object motion, the time evolution of camphor surface concentration, and water flows that transport camphor molecules, which represents a numerically complex problem~\cite{Lauga_Davis_2012,PhysRevFluids.5.084004,PhysRevFluids.6.104006,Ender2021,Ender2021-2,Kang2020}
A number of simplifying approximations, such as the reduction of the hydrodynamic transport model to an effective diffusion constant in the equation for the time evolution of surface camphor concentration, helped to speed up the numerical simulations~\cite{10.1063/1.5021502,Suematsu_Langmuir}.

\begin{figure}	
\centering
\includegraphics{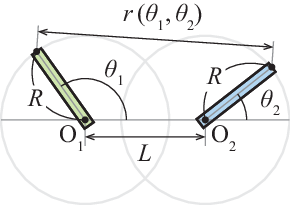}
\caption{The system of two self-propelled camphor-adsorbed ribbons studied in~\cite{PhysRevE.108.024217, PhysRevE.110.064208} and modeled using the Hamiltonian dynamics. The lengths of both ribbons ($R$) were identical. Points $\mathrm{O}_1$ and $\mathrm{O}_2$ at the distance $L$ mark the positions of the axes. A state of the system can be described by the angles $\theta_1$ and $\theta_2$ and 
the angular velocities $\omega_1$ and $\omega_2$. The distance between ribbons ends $r(\theta_1,\theta_2)$ was used to calculate the potential energy.}
\label{fig1} 
\end{figure}

The series of recent publications~\cite{PhysRevE.110.064208, PhysRevE.108.024217, PhysRevE.99.012204, PhysRevE.106.024201, PhysRevE.101.052202, PhysRevE.103.012214, PhysRevE.105.014216} have been concerned with an investigation on the time evolution of camphor-adsorbed self-propelled ribbons that rotate on the water surface around the fixed axes. In few of these papers~\cite{PhysRevE.99.012204, PhysRevE.106.024201, PhysRevE.108.024217, PhysRevE.110.064208}, the authors derived the time evolution equations assuming that the system is described by a Hamiltonian, including the rotational energy of ribbons and the potential energy of their interactions. For example, a state of the system of two ribbons rotating on a water surface illustrated in Fig.~\ref{fig1} can be described by the angles $\theta_1$ and $\theta_2$ describing the ribbon orientations and the angular velocities $\omega_1$ and $\omega_2$.
The Hamiltonian of this system can be written as \cite{PhysRevE.108.024217, PhysRevE.110.064208}:
\begin{align}
H(\theta_1,\theta_2, \omega_1, \omega_2) = \frac{I_1}{2} {\omega_1}^2 + \frac{I_2}{2} {\omega_2}^2 + V (r(\theta_1, \theta_2))
\label{eq1}
\end{align}
and the standard approach is used to derive the Hamiltonian equation. In Eq.~\eqref{eq1}, $I_1$ and $I_2$ are the ribbons' moments of inertia and $V(r(\theta_1, \theta_2))$ is the potential of interactions between them. It was assumed to depend on the distance between the ribbons ends $r(\theta_1, \theta_2)$ (cf.~Fig.~\ref{fig1} and also Fig.~2 in ~\cite{PhysRevE.108.024217} and Fig.~1(c) in ~\cite{PhysRevE.110.064208}). Depending on the publication, the authors used ad-hoc selected potentials such as: Yukawa type one $V(r) = \exp(- K r)/r$~\cite{PhysRevE.99.012204, PhysRevE.101.052202, PhysRevE.103.012214, PhysRevE.106.024201, PhysRevE.108.024217} or a slower decaying potentials of the form  $V(r) = (1 - \exp(- K r^n))/r$ where 
$n \in \{1,2,3\}$~\cite{PhysRevE.110.064208}.

\begin{figure}	
\centering
\includegraphics{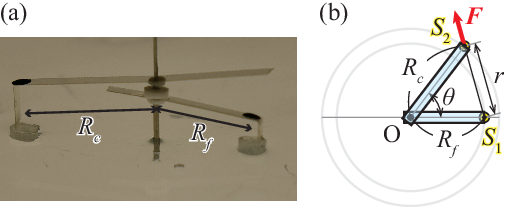}
\caption{The perturbed rotor investigated in our experiments and numerical simulations. (a) Photograph of the experimental setup. The arms were coaxially located. They were lifted above the water surface, and only the camphor disks (diameter: $4$~mm) touched the surface.  The shorter arm ($R_f = 25$~mm) was fixed and did not move. The longer arm ($R_c = 30$~mm) could rotate around the axis. (b) Schematic illustration of the system as seen from the top with definitions of variables used and locations of camphor sources (yellow circles $S_1$ and $S_2$).}
\label{fig2} 
\end{figure}

\section{Results}

The numerical complexity of models based on Hamiltonian equations is significantly lower than that of equations describing object motion and the time evolution of camphor concentration~\cite{doi:10.1021/jp004505n,NAGAYAMA2004151,book-chap2,C5CP00541H,Koyano_PhysRevE.96.012609,Koyano_Chaos_10.1063/1.5061027}. However, one can doubt if such models are correct because the Hamiltonian equations are derived assuming the conservation of system energy. This assumption is clearly not satisfied for self-propelled rotors because the drag forces are continuously acting on the objects moving on the water surface. In order to qualitatively verify whether the Hamiltonian approach introduced in~\cite{PhysRevE.99.012204, PhysRevE.101.052202, PhysRevE.103.012214, PhysRevE.106.024201, PhysRevE.108.024217} leads to a physically correct description, we performed experiments for a simplified set of rotors illustrated in Fig.~\ref{fig2}. We considered two camphor disks at the ends of arms of different lengths. The longer arm can rotate around a vertical axis. Its motion is perturbed by camphor molecules dissipated from a non-moving disk attached at the end of the shorter arm which was fixed.

The Hamiltonian approach and Eq.~\eqref{eq1} can be applied to describe our experiment if we set $\omega_2 \equiv 0$. For the geometry of arms illustrated in Fig.~\ref{fig2}, the distance between the camphor disks is 
\begin{align}
r(\theta)= \sqrt{{R_f}^2 + {R_c}^2 - 2 R_f R_c \cos\theta}.
\label{eq2}
\end{align}
The relationship between the rotor location $\theta$ and the expression for time dependent angular speed $\left|\omega(t)\right|$ strictly derives from energy conservation: 
\begin{align}
\left|\omega(t) \right|= \sqrt{\frac{(2 (E_0 - V(r(\theta(t)))}{I_1}},
\label{eq3}
\end{align}
where $E_0 = I_1 \left\{\omega(t=0)\right\}^2/2 + V(r(\theta(t=0)))$. It follows from Eqs.~\eqref{eq2} and \eqref{eq3} that for any distance-dependent potential and the perturbation located at the $x$-axis, {\bf the angular speed is an even function of the angle $\theta$}.

\begin{figure}	
\centering
\includegraphics[scale=0.95]{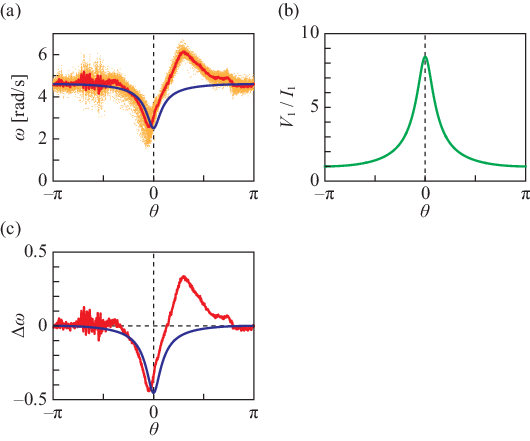}
\caption{
The comparison between experimental results and the Hamiltonian model for a system with a 30-mm-long moving arm and perturbation fixed at 25~mm from the rotation axis. (a) The cumulated experimental data for angular velocity as the function of angle collected for over 300 rotations (orange points), averaged experimental data within $0.01~\mathrm{rad}$ intervals of angle ($\bar{\omega}(\theta)$; the red line), and their fit based on the Hamiltonian model with potential $V_1(r) = \alpha ~{(1-\exp(- \kappa r))/ r}$ where $\alpha/I_1 = 5.5~\mathrm{(rad/s)}^2$ and $\kappa=0.29~\mathrm{mm}^{-1}$ (the blue line). (b)  
The potential selected for the best fit of the experimental angular velocity. (c) The relative change in angular velocity $\Delta\omega (\theta) $ observed in the experiment and predicted by the Hamiltonian model (red and blue lines, respectively).
}
\label{fig3} 
\end{figure}

In our experiments, the diameters of both camphor disks were $4$~mm. The lengths of the shorter and longer arms were $R_f = 25$~mm and $R_c = 30$~mm, respectively (cf.~photograph showing the system geometry in Fig.~\ref{fig2}(a)).
The time evolution of the system state described by the moving rotor orientation $\theta(t)$ and the angular velocity $\omega(t)$ was studied for circa $20$~min. Analyzing experimental results, we digitally recorded the positions of the black dots marking the arm ends and calculated $\omega(t)$ using the ImageJ software~\cite{ImageJ}. During the experiment, camphor molecules continuously evaporated and dissolved in water. Therefore, the experimental conditions changed over time. The results presented in Fig.~\ref{fig3}, illustrate $\omega$ as the function of $\theta$ during a 9-minute-long observation that started 10~min after the experiment was initiated, at which the experimental conditions were stable.
The rotation period was below 2~s; thus, the results corresponded to more than 300~rotations. The cumulated experimental data are illustrated as orange dots in Fig.~\ref{fig3}(a). The angular velocity was always positive, which means that the arm rotated counterclockwise. The angular speed was roughly constant when the arm was away from the perturbation located at $\theta = 0$ and its value was $\omega_0 \sim 4.6$~rad/s. The rotor slowed down when it got closer to the perturbation and its speed dropped below $2$~rad/s. After passing the perturbation at $\theta = 0$, the value of $\omega$ rapidly increased and $\omega(\theta =1~\mathrm{rad})$ was by almost $2~\mathrm{rad/s}$ larger than $\omega_0$. Such a behavior of $\omega(\theta)$ clearly contradicts the prediction of the Hamiltonian model. We considered different potentials suggested in~\cite{PhysRevE.110.064208} and the closest approximation of the experimental data was obtained for $V_1(r) = \alpha~ {(1-\exp(- \kappa r))/ r}$ where $\alpha/I_1 = 5.5~(\mathrm{rad/s})^2$ and $\kappa=0.29 ~\mathrm{mm}^{-1}$, as shown in Fig.~\ref{fig3}(a,c). The values of $\alpha$ and $\kappa$ were adjusted to best match the decay in angular velocity when the rotor approached the perturbation at $\theta = 0$. 
The potential $V_1(\theta)$ along the rotor trajectory is illustrated with a green line in Fig.~\ref{fig3}(b). Equation~\eqref{eq3} gives correct approximation of $\omega(\theta)$ for the rotor approaching the perturbation and completely fails to predict $\omega(\theta)$ for an outgoing rotor.

More precise comparison between experiments and models can be obtained if we average the experimental data. 
The interval of angles $[-\pi,\pi]$ was divided into sub-intervals $0.01~\mathrm{rad}$ long and the value of $\omega(\theta)$ was separately averaged in each sub-interval to give $\bar{\omega}(\theta)$. These results are shown in Fig.~\ref{fig3}(a) (red line) together with the cumulated experimental data. Finally, the average was transformed to calculate the relative change in the average angular velocity with respect to $\omega_0$,
\begin{align}
\Delta\omega (\theta) = \frac{\bar{\omega}(\theta) - \omega_0 }{\omega_0} .
\label{eq4}
\end{align}
The function $\Delta\omega (\theta)$ brings detailed information on the relationship between the decrease in speed when the rotor approaches perturbation at $\theta = 0$ and when it moves away from it. The dependence of $\Delta\omega (\theta)$ on $\theta$ clearly illustrates non-symmetric behavior around the perturbation for the approaching and departing rotor. This asymmetry cannot be predicted by the Hamiltonian model with a distance-dependent potential. The considered experimental data, $\Delta\omega (\theta) $, together with the model prediction, are shown in Fig.~\ref{fig3}(c). The red and blue lines represent the relative change in angular velocity observed in the experiment and obtained in the Hamiltonian model, respectively. The experimentally observed asymmetric $\Delta\omega (\theta)$ with respect to $\theta = 0$ clearly contrasts with the predictions by the Hamiltonian model.

More realistic model for the time evolution of the rotor motion can be formulated if we include the hydrodynamic drag in the evolution equation for the angular velocity. The time evolution equation for the moving arm has the form
\begin{align}
\frac{d\theta}{dt} = \omega(t),
\end{align}
\begin{align}
I_1 \frac{d\omega}{dt} = - \eta (\omega(t) - \omega_0) + \tau_V(\theta(t)),
\label{eq5}
\end{align}
where $\tau_V(\theta(t)) = (R_c R_f \sin \theta /r) (d V/d r)$ is the angle dependent torque generated by interactions with the perturbation and the standard drag term $- \eta (\omega(t) - \omega_0)$ is considered \cite{PhysRevE.101.052202, PhysRevE.103.012214}.

We tested different forms of potential and the values of their parameters to model the changes in $\Delta\omega (\theta) $ around $\theta=0$ observed in experiments. Potentials in the form of $V_1(r)$ did not produce the velocity increase above $\omega_0$ for small $\eta$ for a rotor moving away from the perturbation. Results for $\eta/I_1=8~\mathrm{s}^{-1}$ give the minimum of $\Delta\omega (\theta) \sim -0.015$ on the approach and maximum $\Delta\omega (\theta) \sim 0.005$ when the rotor goes away from perturbation.  

\begin{figure}	
\centering
\includegraphics[scale=0.95]{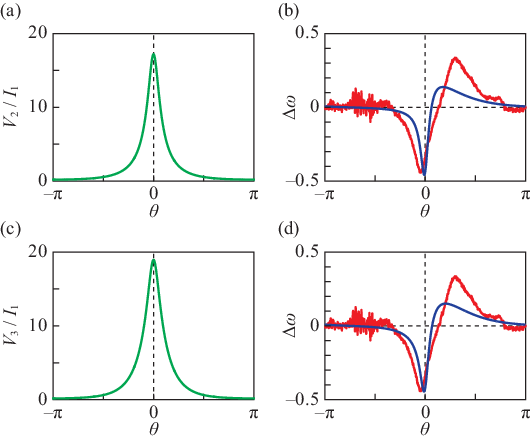}
\caption{Potentials used for the model based on Eq.~\eqref{eq5} (a,c) and the comparison (b,d) between values of $\Delta\omega (\theta)$ observed in the experiment (red line) and calculated using numerical solution of Eq.~\eqref{eq5} (blue line).
(a, b) Potential  $V_2(r) = \alpha~ {\exp(- \kappa r))/ r}$ with with parameter values $\alpha/I_1 = 10.5~(\mathrm{rad/s})^2$, $\kappa = 0.04~\mathrm{mm}^{-1}$ and $\eta/I_1 = 8~\mathrm{s}^{-1}$ in the drag term. (c, d) Potential $V_3(r) = \alpha ~\exp(- \kappa r))/ \sqrt{r}$ with parameters $\alpha/I_1 = 19~(\mathrm{rad/s})^2$, $\kappa = 0.07~\mathrm{mm}^{-1}$ and $\eta/I_1 = 8~\mathrm{s}^{-1}$. 
}
\label{fig4} 
\end{figure}

Much better agreement with experimental results was obtained for potentials that rapidly change close to perturbation. The blue line in Fig.~\ref{fig4}(b) represents the solution of Eq.~\eqref{eq5} for the interaction potential in the form $V_2(r) = \alpha~ {\exp(- \kappa r))/ r}$ with parameter values $\alpha/I_1 = 10.5 ~(\mathrm{rad/s})^2$ and $\kappa = 0.04~\mathrm{mm}^{-1}$. The parameters of the drag term were $\eta/I_1 = 8~\mathrm{s}^{-1}$ and $\omega_0 = 4.6~\mathrm{rad/s}$ that matches experimental data.
The values of potential along the camphor disk trajectory are illustrated in Fig.~\ref{fig4}(a). The angular velocity calculated using Eq.~\eqref{eq5} gives a more realistic qualitative description of the experimental results than the Hamiltonian approach. As in the experimental result, (the red line in Fig.~\ref{fig4}(b)), the minimum speed is shifted from $\theta = 0$ towards negative angles, which means that it is observed before the disk powering the rotor is at the closest distance to the perturbation.  
 
Yet, a better approximation of the experimental results can be obtained for other potential functions. The blue line in Fig.~\ref{fig4}(d) represents the solution of Eq.~\eqref{eq5} for the interaction potential in the form  $V_3(r) = \alpha \exp(- \kappa r))/ \sqrt{r}$ with parameters $\alpha/I_1 = 19~(\mathrm{rad/s})^2$, $\kappa = 0.07~\mathrm{mm}^{-1}$ and $\eta/I_1 = 8~\mathrm{s}^{-1}$ (See Fig.~\ref{fig4}(c)). Such a shape of potential, not considered in the papers we comment on, can be supported by the model based on the coupled equations for the rotor motion and camphor surface concentration. The form of $V_3(r)$ approximates the 2nd-kind modified Bessel function as a stationary solution for the surface concentration of the two-dimensional diffusion-evaporation equation with a disk-shaped source.

\section{Discussion and conclusions}

In the series of recent papers \cite{PhysRevE.99.012204,PhysRevE.106.024201,PhysRevE.108.024217,PhysRevE.110.064208}, the Hamiltonian equations were applied to simulate the time evolution of camphor adsorbed self-propelled ribbons rotating on the water surface. The idea is attractive because if a distance-dependent potential describes the interactions between self-propelled objects, then the model has a low numerical complexity if compared with the description based on the equation of motion combined with reaction-diffusion model for camphor concentration~\cite{book-chap2,NAGAYAMA2004151,Koyano_PhysRevE.96.012609,Koyano_Chaos_10.1063/1.5061027,Koyano_2021PhysRevE,Kitahata_2023PhysRevE}. On the other hand, the Hamiltonian approach assumes the conservation of energy, whereas energy dissipation due to hydrodynamic drag can be clearly seen in experiments. As a test we applied it to the description of a simple experiment in which the motion of a single camphor rotor was perturbed by a camphor disk fixed on the water surface. The analysis of the rotor motion shows that the Hamiltonian model is wrong because it does not explain the asymmetry between the rotor's angular velocity observed in experiments when it approaches or moves away from the perturbation.
If the method fails to describe a single rotor, we doubt it can correctly treat many rotor systems.
In our opinion, this is a serious argument against the application of Hamiltonian approach to self-propelled camphor objects.

We have also demonstrated that the idea of interaction potential can give qualitatively correct behavior of $\Delta\omega (\theta) $ if the equation of motion includes the hydrodynamic drag. However, authors in  \cite{PhysRevE.99.012204,PhysRevE.101.052202,PhysRevE.103.012214,PhysRevE.110.064208} did not describe how the interaction potentials for specific systems were selected. They just used potentials of simple forms in their simulations. 
We demonstrated that our experimental results could be reasonably well fitted with  $V_3(r) = \alpha \exp(- \kappa r)/ \sqrt{r}$ that, as we discuss below, can be related to a stationary solution of the diffusion-evaporation problem, i.e., the 2nd-kind modified Bessel function. 

Marangoni forces act on objects floating on the water surface because of gradients in surface tension. In the considered system, the surface is covered with an unstable camphor layer characterized by time-dependent surface concentration $u(\bm{q},t)$, where $\bm{q} $ is the surface coordinate. The decrease in water surface tension $\gamma(\bm{q},t)$ at the point where the surface camphor concentration is $u(\bm{q},t)$ can be approximated by a linear decreasing function:
\begin{align}
\gamma(\bm{q},t) = \gamma_0 - \Gamma u(\bm{q},t)
\label{eq6}
\end{align}
In Eq.~\eqref{eq6}, $\Gamma >0$ and $\gamma_0$ is the surface tension of pure water.

The resulting force $\bm{F}$ acting on an object can be calculated as \cite{book-chap2}:
\begin{align}
    \bm{F} = \iint_A \left(\nabla \gamma \right) \chi (\bm{q},t) d\bm{q}  = - \Gamma \iint_A \left(\nabla u \right) \chi (\bm{q},t) d\bm{q},
\label{eq7}
\end{align}
where integration is performed over the whole water surface $A$ and $\chi (\bm{q},t)$ is the time dependent characteristic function of the object.
Knowing $\bm{F}$, we can calculate the torque acting on the rotor and study its time evolution.

To describe our experimental system, we considered $A$ is a disk with the radius $R_\mathrm{sys}$. The Cartesian coordinates were $\bm{q} = x \bm{e}_x + y \bm{e}_y$, with the unit vectors $\bm{e}_x$ and $\bm{e}_y$ oriented in  positive $x$ and $y$ directions. The origin of the system as well as the rotor axis were set at the center of $A$. The rotor arm length was $R_c$ as illustrated in Fig.~\ref{fig2}(b). Using $\theta(t)$ as the state variable, the center position of the rotor propelling camphor disk is $\bm{q}_c(t) = R_c\left( \bm{e}_x\cos\theta(t)  + \bm{e}_y \sin \theta(t) \right)$, and the center position of the fixed camphor disk is at $\bm{q}_f = R_f \bm{e}_x$.

The motion of the rotor is given by the Newtonian equation as:
\begin{align}
I_1 \frac{d^2\theta}{dt^2}= - \mu {R_c}^2 \frac{d\theta}{dt} + \bm{q}_c \times \bm{F},
\label{eq8}
\end{align}
where $\mu$ is the coefficient of hydrodynamic drag, and the evolution of $u(\bm{q},t)$ is described by
\begin{align}
    \frac{\partial u}{\partial t} = \nabla^2 u - u + S(\bm{q}; \bm{q}_c(t), \bm{q}_f).
\label{eq9}
\end{align}
The first term in Eq.~\eqref{eq9} corresponds to the effective diffusion of the camphor at the water surface that reflects not only the thermal diffusion but also the Marangoni convection~\cite{10.1063/1.5021502,Suematsu_Langmuir}. The second term corresponds to camphor evaporation to the air phase and dissolution to the bulk phase. The third term $S(\bm{q}; \bm{q}_c(t), \bm{q}_f)$ describes the inflow of camphor from both rotating and fixed camphor disks. It can be decomposed as:
\begin{align}
    S(\bm{q}; \bm{q}_c(t),\bm{q}_f) = \chi_c(\bm{q}; \bm{q}_c(t)) +  \chi_f(\bm{q}; \bm{q}_f),
\label{eq10}
\end{align}
where $\chi_c(\bm{q}; \bm{q}_c(t))$ and $\chi_f(\bm{q}; \bm{q}_f) $ are characteristic functions of rotating and fixed disks. 
The first and second terms in the right side denote the supply from the camphor disks propelling the rotor and the fixed one, respectively. 

The units of dimensionless variables are defined as follows: The distance unit is a diffusion length $\sqrt{D/a}$, where $D$ is the effective diffusion coefficient and $a$ is the rate of evaporation and dissolution. The time unit is given by $1/a$. The unit of force is defined assuming that the coefficient $\Gamma = 1$. Finally, the concentration unit is determined by $S_0/a$, where $S_0$ is the supply rate from a camphor disk. Comparing with the experiment, the space and time units are roughly estimated as $10~\mathrm{mm}$ and $1~\mathrm{s}$.

To avoid numerical numerical instabilities while solving the time evolution equation for $u(\bm{q},t)$ (Eq.~\eqref{eq9}), the characteristic functions were approximated by a smooth function, which is almost $1$ inside a camphor disk and almost $0$ outside it:
\begin{align}
    \chi_c(\bm{q}; \bm{q}_c(t)) = \frac{1}{2}\left[1 + \tanh \left( \frac{(d/2)
- \left| \bm{q} - \bm{q}_c(t)\right|}{\delta} \right)\right],
\end{align}
\begin{align}
    \chi_f(\bm{q}; \bm{q}_f) = \frac{1}{2}\left[1 + \tanh \left( \frac{(d/2)
- \left| \bm{q} - \bm{q}_f\right|}{\delta} \right)\right],
\end{align}
where $d$ is the camphor disk diameter and $\delta$ is the characteristic length for the smoothing.

The numerical simulation was performed with the Euler method with a time step of $\Delta t =0.0001$ and the spatial mesh $\Delta q = 0.025$. The parameters are set as $I_1 = 9 \times 10^{-5}$, $R_c = 3$, $R_f = 2.5$, $d = 0.4$, $R_\mathrm{sys} = 9$, $\mu = 0.00235$, and $\delta = 0.025$.

The model shows that when the fixed camphor disk is absent, i.e.,
\begin{align}
    S(\bm{q}; \bm{q}_c (t),\bm{q}_f) = \chi_c(\bm{q}; \bm{q}_c(t)) , 
\end{align}
the arm rotated with a constant angular velocity for the smaller $\mu$, while it stops and does not show rotation for the greater $\mu$. The transition between still arm and rotation occurred at $\mu_c \cong 0.015$. Here, we set $\mu = 0.00235$, at which the stable angular velocity  $\omega_0 = 4.6004$.

\begin{figure}
\centering
\includegraphics[scale=0.95]{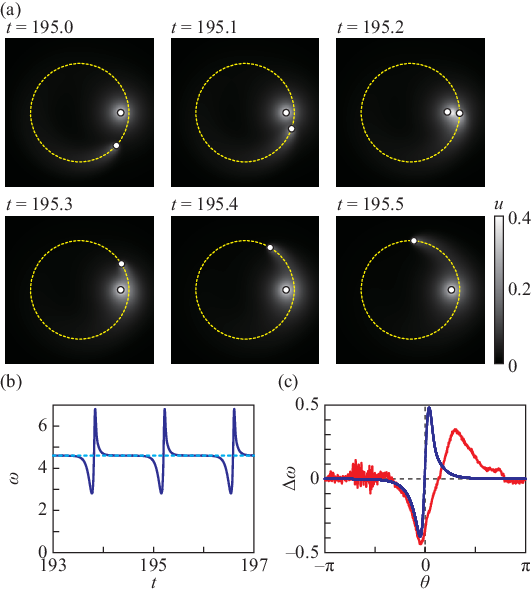}
\caption{ Results of numerical simulation based on the reaction-diffusion equation coupled with the Newtonian equation for the rotor. (a) Snapshots of the camphor concentration field and the position of the camphor disk at $t=195.0$, $195.1$, $195.2$, $195.3$, $195.4$, and $195.5$. The brighter region indicates the higher camphor concentration. The yellow circle corresponds to the trajectory of the camphor disk. (b) Time series of the angular velocity $\omega$ of the rotor (blue thick curve). The cyan dotted line shows the angular velocity $\omega_0$, which is the converged angular velocity of the rotor without another fixed camphor source. (c) Normalized angular velocity $\Delta \omega (\theta)$: blue and red lines correspond to the numerical and experimental results, respectively.}
\label{fig5}
\end{figure}

The results of numerical simulations with a perturbing camphor disk are presented in Fig.~\ref{fig5}. In Fig.~\ref{fig5}(a), the snapshots made every time step $0.1$ illustrate the asymmetric camphor concentration field is between the front and rear of the camphor rotor. In Fig.~\ref{fig5}(b), the  angular velocity $\omega$ as the function of time is presented. It clearly shows the periodic deceleration followed by acceleration, which is the effect by the concentration field generated by the fixed camphor disk. The normalized angular velocity as the function of rotor angle $\theta$ $\Delta\omega (\theta) $ is plotted in Fig.~\ref{fig5}(c). As seen in Fig. \ref{fig5} the model based on Eqs.~\eqref{eq8} and \eqref{eq9} and the parameters matching experimental conditions gives as good description of the experimental results as that with the optimized potential.

Finally, let us discuss the relationship between the potential model of rotor motion (Eq.~\eqref{eq5}) and the coupled evolution equations for the object location and surface concentration of camphor. Since both Eqs.~\eqref{eq8} and \eqref{eq9} are linear, the solution of Eq.~\eqref{eq9} can be written as:
\begin{align}
u(\bm{q},t) = u_c(\bm{q},t) + u_f(\bm{q},t),
\end{align}
where  $u_c(\bm{q},t) $ and $ u_f(\bm{q},t)$ are solutions of equations similar to Eq.~\eqref{eq9} but with $\chi_c(\bm{q}; \bm{q}_c(t))$ and $\chi_f(\bm{q}; \bm{q}_f)$ instead of $ S(\bm{q}; \bm{q}_c(t),\bm{q}_f)$. Now, the time evolution equation has the form:
\begin{align}
I_1 \frac{d^2\theta}{dt^2}=& - \mu {R_c}^2 \frac{d\theta}{dt} + \Gamma \bm{q}_c \times \iint_A \left(\nabla u_c \right) \chi_c(\bm{q}; \bm{q}_c(t)) d\bm{q} \nonumber \\
& + \Gamma \bm{q}_c \times \iint_A \left(\nabla u_f \right) \chi_c(\bm{q}; \bm{q}_c(t)) d\bm{q}.
\label{eq14}
\end{align}
Assuming that, in the coordinate system related to the camphor disk on the rotating arm, the surface concentration $u_c(\bm{q},t)$ does not depend on time, and it is the same as that for stationary moving unperturbed arm, we have:
\begin{align}
\mu R_c^2 \omega_0  =  \Gamma \bm{q}_c \times \iint_A \left(\nabla u_c \right) \chi_c(\bm{q}; \bm{q}_c(t)) d\bm{q}.
\label{eq15}
\end{align}
Thus, the time evolution equation has the form:
\begin{align}
I_1 \frac{d\omega}{dt} =& - \mu R_c^2 (\omega - \omega_0) \nonumber \\
& + \Gamma \bm{q}_c \times \iint_A \left(\nabla u_f \right) \chi_c(\bm{q}; \bm{q}_c(t)) d\bm{q}.
\end{align}
If the moving object is small, then
\begin{align}
\iint_A \left(\nabla u_f \right) \chi_c(\bm{q}; \bm{q}_c(t)) d\bm{q} \cong \Omega ~\nabla u_f |_{\bm{q}_c},
\label{eq17}
\end{align}
where $\Omega$ is the rotating disk surface.
It means that within the approximations introduced, the camphor concentration profile generated by a fixed disk plays the same role as the interaction  potential~\cite{PhysRevE.110.064208, PhysRevE.108.024217, PhysRevE.99.012204, PhysRevE.106.024201, PhysRevE.101.052202, PhysRevE.103.012214, PhysRevE.105.014216} because the force can be calculated as its gradient. This observation justifies the application of potential in the form $V_3(r)$ that is a fair approximation of the solution of Eq.~\eqref{eq9} for camphor concentration profile around a disk-shaped source and hopefully can be generalized to other systems, too. In the case of complex shape of propelled object the integration in Eq. \eqref{eq17} can be replaced by an integration quadrature, that leads to "interaction potential" in a many-body form.

\begin{acknowledgments}
    This work was supported by PAN-JSPS program ``Complexity and order in systems of deformable self-propelled objects'' (No.~JPJSBP120234601). 
\end{acknowledgments}



%
\end{document}